\documentclass[prb,superscriptaddress,showpacs,floatfix,onecolumn]{revtex4}
\usepackage{amssymb}
\usepackage{amsmath}
\usepackage{graphicx}

\begin{document}

\title{Replica study of pinned bubble crystals }
\author{R. C\^{o}t\'{e}}
\affiliation{D\'{e}partement de physique and RQMP, Universit\'{e} de Sherbrooke,
Sherbrooke, Qu\'{e}bec, Canada, J1K 2R1}
\author{Mei-Rong Li}
\affiliation{D\'{e}partement de physique and RQMP, Universit\'{e} de Sherbrooke,
Sherbrooke, Qu\'{e}bec, Canada, J1K 2R1}
\author{A. Faribault}
\affiliation{D\'{e}partement de physique and RQMP, Universit\'{e} de Sherbrooke,
Sherbrooke, Qu\'{e}bec, Canada, J1K 2R1}
\author{H. A. Fertig}
\affiliation{Department of Physics, Indiana University, Bloomington, Indiana 47405}

\begin{abstract}
In higher Landau levels ($N>1$), the ground state of the two-dimensional
electron gas in a strong perpendicular magnetic field evolves from a Wigner
crystal for small filling $\nu $ of the partially filled Landau level, into
a succession of bubble states with increasing number of guiding centers per
bubble as $\nu $ increases, to a modulated stripe state near $\nu =0.5$. In
this work, we compute the frequency-dependent longitudinal conductivity $%
\sigma _{xx}\left( \omega \right) $ of the Wigner and bubble crystal states
in the presence of disorder. We apply an elastic theory to the crystal
states which is characterized by a shear and a bulk modulus. We obtain both
moduli from the microscopic time-dependent Hartree-Fock approximation. We
then use the replica and Gaussian variational methods to handle the effects
of disorder. Within the semiclassical approximation we get the dynamical
conductivity as well as the pinning frequency as functions of the Landau
level filling factor and compare our results with recent microwave
experiments.
\end{abstract}

\date{\today}
\maketitle

\section{Introduction}

In the presence of a strong quantizing magnetic field, the two-dimensional
electron gas (2DEG) is expected to crystallize below a filling factor $\nu
\sim 1/6.5.$\cite{lamgirvin} The resulting electron solid, or Wigner crystal
(WC), has been intensively studied, theoretically, since the early days of
the quantum Hall effect. On the experimental side, many groups have reported
the observation of a strong increase in the diagonal resistivity $\rho _{xx}$
together with non-linear I-V characteristics and broadband noise when the
filling factor of the lowest Landau level is decreased below $1/5.$ These
observations have been interpreted as the pinning and sliding of a Wigner
crystal.\cite{reviewexperimentwc} Early microwave absorption experiments\cite%
{ye} detected a resonance in the real part of the longitudinal conductivity
tensor, $\sigma _{xx}\left( \omega \right) ,$ that was attributed to the
formation of a pinned Wigner crystal. (A more recent experiment\cite{chen2}
reports the observation of not one but two resonances at low filling factor.
The origin of these resonances are not yet fully understood.) More recent
studies show a resonance in the absorption at filling factors close to $\nu
=1,2,3$ where the formation of a Wigner solid is expected in very clean
samples.\cite{chen1,lewis1}

In this paper, we concern ourselves with a series of microwave absorption
experiments\cite{lewis2} on the so-called bubble phases in Landau levels $%
N=2.$ The bubble phases, first introduced by Fogler and Koulakov,\cite%
{foglerkoulakov} are basically crystals made of clusters of $M$ electrons.
More precisely, the guiding centers of the cyclotron motion (the cyclotron
radius is given by $R_{c}=\sqrt{2N+1}\ell $ where $\ell =\sqrt{\hslash c/eB}$
is the magnetic length) of the $M$ electrons are concentrated in a small
area of radius $R_{M}=$ $\sqrt{2M}\ell $ at each lattice site. In the
Hartree-Fock approximation, the 2DEG ground state, in Landau levels $N>1$,
evolves from a Wigner crystal ($M=1$), near integer filling, into bubble
crystals with $M=2,...,N+1$ as the filling factor increases, and terminates
with the quantum Hall stripe state near half-filling of the partially filled
level\cite{cotebubble}. (The sequence repeats in inverse order for holes for
partial filling factors between $0.5$ and $1.0$.) All these transitions are
believed to be first order.

Microwave absorption experiments \cite{lewis2} reveal a resonance in the
second Landau level ($N=2$) near integer filling, which was attributed to
the formation of a Wigner crystal. The frequency of this resonance
decreases, as the filling factor $\nu =\nu _{tot}-2N$ of the partially
filled level increases, until it reaches a plateau at $\nu =\nu _{1}\approx
0.16$ where a second resonance appears (see Fig. 3 of Ref. %
\onlinecite{lewis2}). This second resonance is attributed to the formation
of a bubble crystal with $M=2$. The frequencies of both resonances are
nearly constant between $\nu _{1}$ and $\nu _{2}\approx 0.26.$ At $\nu _{2}$%
, the first resonance disappears and the frequency of the second one starts
to decrease. It has been suggested that the presence of the two resonances
between $\nu _{1}$ and $\nu _{2}$ can be explained by the coexistence of the
two crystal phases in this range of filling factor.\cite{dorsey}

In a recent paper, we have studied the dynamics of the bubble crystals in
the absence of disorder\cite{cotebubble}. We have shown that these crystals
have a low-energy magnetophonon mode with dispersion $\omega \sim k^{3/2},$
at small wavevector, that is typical of a pure Wigner crystal in a strong
magnetic field. In the presence of disorder, this low-energy magnetophonon
mode becomes gapped, with lowest energy at the pinning frequency. In this
paper, we compute this pinning frequency and the evolution of the
longitudinal part of the dynamical conductivity, $\sigma _{xx}\left( \omega
\right) ,$ as the 2DEG makes transitions from the WC to the bubble states
and compare our results with those of Lewis et al.\cite{lewis2} (We have
previously addressed the dynamics of the disordered stripe state in Refs. %
\onlinecite{meirong1} and \onlinecite{meirong2}). The pinning frequency is
extracted from the real part of the dynamical conductivity \textit{i.e}.
from $\Re \left[ \sigma _{xx}\left( \mathbf{q}=0,\omega \right) \right] $
computed in the presence of disorder. To deal with the disorder averaging,
we use a combination of the replica trick and Gaussian variational methods
(GVM).\cite{replica} A similar technique has been used by Chitra \emph{et al.%
} \cite{chitra} to compute the dynamical conductivity tensor of the Wigner
crystal in the lowest Landau level. While this method appears to give
reasonable results for this and other pinned systems, some controversy has
arisen regarding the width of the resulting resonant peak for a pinned
Wigner crystal in a magnetic field \cite{fertig99, narrowpeak}. We address
this issue toward the end of this paper.

To deal with the bubble crystals, some modifications of the replica method
are needed. For example, electrons and bubbles are more extended objects in
higher Landau levels and so it is necessary to include a form factor in the
calculation of the dynamical matrix that enters the elastic action as well
as the effective disorder potential. We find that the evolution of the bulk
and shear moduli with filling factor is essentially driven by the changes in
the form factor of the bubbles. A softening of the lattice appears at the
critical filling $\nu _{1}$ introduced above when the outer radii of
adjacent bubbles touch one another and, as we will show, is correctly
captured by a dynamical matrix computed with the correct form factor.
Moreover, the gap between the two pinning resonances seen in the microwave
experiment may be understood as being partially due to the change in form
factor in the transition from the Wigner to the $M=2$ bubble crystal. In our
calculation, the bulk and shear moduli are extracted from the
density-density response function computed in the Generalized-Random-Phase
Approximation (GRPA).\cite{cote2}

Our paper is organized in the following way. In Sec. II, we build an elastic
model for the description of the pure crystal states and extract the
dynamical matrix from the GRPA calculation. In Sec. III, we briefly review
the replica and GVM that we use to handle the disorder. Our numerical
results for the dynamical conductivity and pinning peak are presented in
Sec. IV. We compare our results with the experimental data in Sec. V.
Section VI is devoted to a discussion of the width of the resonance obtained
in this method as compared to others, and we conclude with a summary in Sec.
VII. A brief account of this work has appeared previously.\cite{florida}

\section{ Elastic model for the bubble crystals}

\subsection{Form factor}

In describing the bubble crystals in higher Landau levels, we assume, as
usual, that the electrons in the filled levels are inert. The electrons in
the partially filled Landau level $N$ are assumed to form a triangular
crystal structure\cite{cotebubble} with $M$ electrons associated with each
lattice site. It is convenient to describe the crystal by the Fourier
components $\left\langle \rho \left( \mathbf{K}\right) \right\rangle $ of
the guiding-center density where $\mathbf{K}$ is a reciprocal lattice vector
of the triangular lattice. The guiding center density is related to the real
electronic density\cite{cotebubble} by the relation 
\begin{equation}
\left\langle n\left( \mathbf{K}\right) \right\rangle =\int d\mathbf{r}e^{-i%
\mathbf{K}\cdot \mathbf{r}}\left\langle n\left( \mathbf{r}\right)
\right\rangle =N_{\varphi }F_{N}\left( \mathbf{K}\right) \left\langle \rho
\left( \mathbf{K}\right) \right\rangle ,  \label{a_1}
\end{equation}%
where $\left\langle n\left( \mathbf{K}\right) \right\rangle $ is a Fourier
component of the real density, $N_{\varphi }$ is the Landau-level
degeneracy, and 
\begin{equation}
F_{N}\left( \mathbf{K}\right) =e^{-K^{2}\ell ^{2}/4}L_{N}^{0}\left( \frac{%
K^{2}\ell ^{2}}{2}\right)
\end{equation}%
( $L_{N}^{0}\left( x\right) $ is a generalized Laguerre polynomial) is the
form factor of an electron in Landau level $N$.

\begin{figure}[tbp]
\includegraphics[width=8cm]{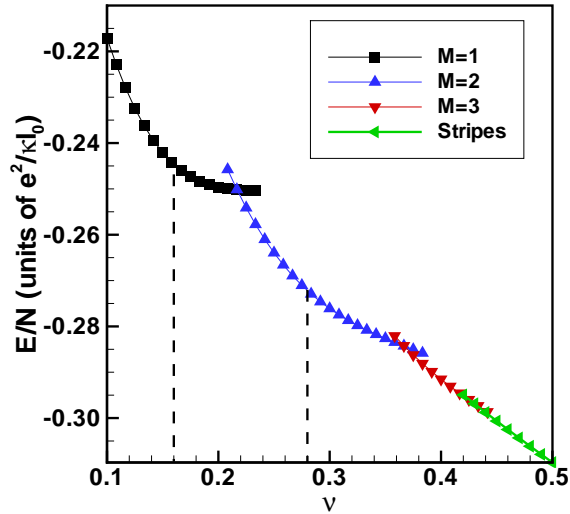}
\caption{Energy per particle for the bubble and stripe states in Landau
level $N=2$. The region between the dashed lines is where the two resonances
are observed to coexist in the microwave experiment.}
\label{fig_energien2}
\end{figure}

The Hartree-Fock energy per electron in the partially filled Landau level is
given by%
\begin{equation}
\frac{E}{N}=\frac{1}{2\nu }\sum_{\mathbf{K}}\left[ H\left( \mathbf{K}\right)
\left( 1-\delta _{\mathbf{K},0}\right) -X\left( \mathbf{K}\right) \right]
\left\vert \left\langle \rho \left( \mathbf{K}\right) \right\rangle
\right\vert ^{2},  \label{a_2}
\end{equation}%
where $\nu =\nu _{tot}-2N$ is the filling factor of the partially filled
level. The Hartree and Fock interactions in Landau level $N$ are given by 
\begin{eqnarray}
H_{N}\left( \mathbf{q}\right) &=&\left( \frac{e^{2}}{\kappa \ell }\right) 
\frac{1}{q\ell }e^{\frac{-q^{2}\ell ^{2}}{2}}\left[ L_{N}^{0}\left( \frac{%
q^{2}\ell ^{2}}{2}\right) \right] ^{2}, \\
X_{N}\left( \mathbf{q}\right) &=&\left( \frac{e^{2}}{\kappa \ell }\right) 
\sqrt{2}\int_{0}^{\infty }dx\,e^{-x^{2}}\left[ L_{N}^{0}\left( x^{2}\right) %
\right] ^{2}J_{0}\left( \sqrt{2}xq\ell \right) ,
\end{eqnarray}%
where $\kappa $ is the dielectric constant of the host material. Figure \ref%
{fig_energien2} shows the Hartree-Fock energy per particle for the bubble
crystals in $N=2$. The first order transition from the WC to the $M=2$
bubble phase occurs at $\nu _{1}=0.22,$ that between the $M=2$ and $M=3$
bubble phases occurs at $\nu _{2}=0.37$ and that to the stripe phase occurs
at $\nu _{3}=0.43.$ In Fig. \ref{fig_energien2}, the energy is in units of $%
e^{2}/\kappa \ell _{0}$ where $\ell _{0}$ is the magnetic length evaluated
at filling factor $\nu _{0}=0.1$ using 
\begin{equation}
\frac{e^{2}}{\kappa \ell }=\sqrt{\frac{2N+\nu _{0}}{2N+\nu }}\left( \frac{%
e^{2}}{\kappa \ell _{0}}\right) ,
\end{equation}%
an expression that takes into account the filled Landau levels. It is
important to notice that the variation of the magnetic length $\ell $
through the $M=1$ or $M=2$ bubble phases is quite small for $N=2$ so that we
can effectively consider the variation of the filling factor to be
equivalent to a variation of the density of electrons. In the microwave
experiments\cite{lewis2}, two resonances are observed in the region bounded
by the dashed lines in Fig.~\ref{fig_energien2}, which is argued to be the
signature of the coexistence of the two phases.

Once the $\left\langle \rho \left( \mathbf{K}\right) \right\rangle ^{\prime
}s$ are known in the HFA, it is possible to compute the Matsubara
two-particle Green's function 
\begin{equation}
\chi _{\mathbf{K},\mathbf{K}^{\prime }}^{\left( \rho ,\rho \right) }\left( 
\mathbf{k},\tau \right) =-N_{\varphi }\left\langle {\mathcal{T}}_{\tau }%
\widetilde{\rho }\left( \mathbf{k}+\mathbf{K},\tau \right) \widetilde{\rho }%
\left( -\mathbf{k}-\mathbf{K}^{\prime },0\right) \right\rangle ,  \label{a_3}
\end{equation}%
in the GRPA\cite{cote2}. In Eq. (\ref{a_3}), $\widetilde{\rho }\equiv \rho
-\left\langle \rho \right\rangle ,$ $\mathbf{k}$ is a wavevector in the
first Brillouin zone of the triangular lattice and ${\mathcal{T}}_{\tau }$
is the time-ordering operator. The analytical continuation $\chi _{\mathbf{K}%
,\mathbf{K}^{\prime }}^{\left( \rho ,\rho \right) }\left( \mathbf{k},\omega
+i\delta \right) $ of this Green's function gives the density response
function whose poles are the collective (density) excitations of the
crystal. The lowest-energy collective mode is the magnetophonon mode. We can
get its dispersion relation, $\omega _{GRPA}\left( \mathbf{k}\right) $ by
following the lowest-energy pole with non-vanishing weight of $\chi _{%
\mathbf{K=0},\mathbf{K}^{\prime }\mathbf{=0}}^{\left( \rho ,\rho \right)
}\left( \mathbf{k},\omega +i\delta \right) $ as $\mathbf{k}$ is varied in
the Brillouin zone.

In order to compute the dynamical conductivity $\sigma _{xx}\left( \omega
\right) $ of the disordered bubble crystals using the replica and GVM, it is
convenient to construct an elastic theory valid for the low-energy
excitations. As has been demonstrated in Refs. \onlinecite{meirong1,meirong2}%
, we can do so by proceeding in the following way. We model the crystals as
being made of rigid extended objects centered on each lattice site $\mathbf{R%
}$ with a density profile (or form factor) given by $h\left( \mathbf{r}%
\right) $\textbf{.} Because each bubble has $M$ electrons, we have the
normalisation%
\begin{equation}
\int d\mathbf{r}h\left( \mathbf{r}\right) =M.
\end{equation}%
The dynamics of these objects is described by the displacement fields $%
\mathbf{u}\left( \mathbf{R},t\right) $ at each lattice site $\mathbf{R}$.
The electronic density can therefore be given by%
\begin{equation}
n\left( \mathbf{r},t\right) =\sum_{\mathbf{R}}h\left( \mathbf{r}-\mathbf{R}-%
\mathbf{u}\left( \mathbf{R},t\right) \right) .  \label{a_4}
\end{equation}%
In the ground state, the HF\ energy given by Eq. (\ref{a_2}) is very well
reproduced if we use the form factors\cite{foglerkoulakov},%
\begin{equation}
h\left( \mathbf{q}\right) =\int d\mathbf{r}e^{-i\mathbf{q}\cdot \mathbf{r}%
}h\left( \mathbf{r}\right) =e^{-q^{2}\ell ^{2}/2},  \label{a_5}
\end{equation}%
for the WC in $N=0$ and by 
\begin{equation}
h\left( \mathbf{q}\right) =\left\{ 
\begin{array}{ccc}
e^{-q^{2}\ell ^{2}/2}\left( 1-\left( q\ell \right) ^{2}+\frac{1}{8}\left(
q\ell \right) ^{4}\right) , & \text{for} & M=1 \\ 
e^{-q^{2}\ell ^{2}/2}\left( 2-\frac{1}{2}\left( q\ell \right) ^{2}\right)
\left( 1-\left( q\ell \right) ^{2}+\frac{1}{8}\left( q\ell \right)
^{4}\right) , & \text{for} & M=2%
\end{array}%
\right.  \label{a_6}
\end{equation}%
for the bubble crystals in $N=2$. These form factors are obtained from the
Slater determinant for the wavefunction of a bubble of $M$ electrons in
Landau level $N$ 
\begin{equation}
\Psi _{N}\left( \mathbf{r}_{1},\mathbf{r}_{2},...,\mathbf{r}_{M}\right)
=\left\vert 
\begin{array}{cccc}
\varphi _{N,0}\left( \mathbf{r}_{1}\right) & \varphi _{N,0}\left( \mathbf{r}%
_{2}\right) & ... & \varphi _{N,0}\left( \mathbf{r}_{M}\right) \\ 
\varphi _{N,1}\left( \mathbf{r}_{1}\right) & \varphi _{N,1}\left( \mathbf{r}%
_{2}\right) & ... & \varphi _{N,1}\left( \mathbf{r}_{M}\right) \\ 
\vdots & \vdots & \vdots & \vdots \\ 
\varphi _{N,M-1}\left( \mathbf{r}_{1}\right) & \varphi _{N,M-1}\left( 
\mathbf{r}_{2}\right) & ... & \varphi _{N,M-1}\left( \mathbf{r}_{M}\right)%
\end{array}%
\right\vert ,
\end{equation}%
where $\varphi _{N,m}\left( \mathbf{r}\right) =C_{N,m}\left( \frac{r}{\ell }%
\right) ^{\left\vert m-N\right\vert }e^{-r^{2}/4\ell ^{2}}e^{i\left(
N-m\right) \theta }L_{\left( N+m-\left\vert m-N\right\vert \right)
/2}^{\left\vert m-N\right\vert }\left( \frac{r^{2}}{2\ell ^{2}}\right) $ is
the normalized wave function of an electron in the symmetric gauge $\mathbf{A%
}=\left( -B_{0}y/2,B_{0}x/2,0\right) $ with angular momentum $m$. The
associated one-particle density is simply%
\begin{equation}
h\left( \mathbf{r}\right) =\left[ \prod_{i=2}^{M}\int d\mathbf{r}_{i}\right]
\left\vert \Psi _{N}\left( \mathbf{r},\mathbf{r}_{2},...,\mathbf{r}%
_{M}\right) \right\vert ^{2}=\sum_{m=0}^{m=M-1}\left\vert \varphi
_{N,m}\left( \mathbf{r}\right) \right\vert ^{2}.
\end{equation}

In the presence of a transverse magnetic field $\mathbf{B}=-B_{0}\widehat{%
\mathbf{z}}$, the equation of motion for the displacement field of the rigid
object of charge $Me$ and mass $Mm^{\ast }$ on site $i$ is given by%
\begin{equation}
\left( Mm^{\ast }\right) \frac{d^{2}\mathbf{u}\left( \mathbf{R}_{i}\mathbf{,}%
t\right) }{dt^{2}}=\frac{MeB_{0}}{c}\frac{d\mathbf{u}\left( \mathbf{R}_{i},%
\mathbf{\,}t\right) \times \widehat{\mathbf{z}}}{dt}-\sum_{j}\widehat{D}%
\left( \mathbf{R}_{i}-\mathbf{R}_{j}\right) \cdot \mathbf{u}\left( \mathbf{R}%
_{j},t\right) ,  \label{a_7}
\end{equation}%
where $\widehat{D}\left( \mathbf{R}_{i}-\mathbf{R}_{j}\right) $ is the
dynamical matrix of the crystal (we use a \textquotedblleft
hat\textquotedblright\ to indicate that a quantity is a $2\times 2$ matrix).
The dynamics in the strong-magnetic field limit is obtained by setting $%
m^{\ast }=0$ in Eq. (\ref{a_7}) so that, after Fourier transforming, we get 
\begin{equation}
\frac{M\hslash }{\ell ^{2}}\frac{d\mathbf{u}\left( \mathbf{k},t\right) }{dt}%
\times \widehat{\mathbf{z}}-\widehat{D}\left( \mathbf{k}\right) \cdot 
\mathbf{u}\left( \mathbf{k},t\right) =0,
\end{equation}%
where we have defined%
\begin{equation}
\widehat{D}\left( \mathbf{R}_{i}-\mathbf{R}_{j}\right) =\frac{1}{N_{s}}\sum_{%
\mathbf{k}}\widehat{D}\left( \mathbf{k}\right) e^{i\mathbf{k}\cdot \left( 
\mathbf{R}_{i}-\mathbf{R}_{j}\right) },
\end{equation}%
and%
\begin{equation}
\mathbf{u}\left( \mathbf{k},t\right) =\frac{1}{\sqrt{N_{s}}}\sum_{\mathbf{R}%
}e^{-i\mathbf{k}\cdot \mathbf{R}}\mathbf{u}\left( \mathbf{R},t\right) ,
\end{equation}%
with $N_{s}$ the number of lattice sites. Fourier transforming in time $%
\mathbf{u}\left( \mathbf{k},t\right) =\int \frac{d\omega }{2\pi }e^{i\omega
t}\mathbf{u}\left( \mathbf{k},\omega \right) ,$ we have, in matrix form 
\begin{equation}
\left( 
\begin{array}{cc}
D_{xx}\left( \mathbf{k}\right) & D_{xy}\left( \mathbf{k}\right) -i\frac{%
\hslash M}{\ell ^{2}}\omega \\ 
D_{yx}\left( \mathbf{k}\right) +i\frac{\hslash M}{\ell ^{2}}\omega & 
D_{yy}\left( \mathbf{k}\right)%
\end{array}%
\right) \left( 
\begin{array}{c}
u_{x}\left( \mathbf{k},\omega \right) \\ 
u_{y}\left( \mathbf{k},\omega \right)%
\end{array}%
\right) =\left( 
\begin{array}{c}
0 \\ 
0%
\end{array}%
\right) .  \label{a_8}
\end{equation}%
Using Eq. (\ref{a_8}), we get for the magnetophonon dispersion relation%
\begin{equation}
\omega =\frac{\ell ^{2}}{\hslash M}\sqrt{\det \left[ \widehat{D}\left( 
\mathbf{k}\right) \right] }.  \label{a_9}
\end{equation}

\subsection{Euclidean Elastic Action}

Using Eq. (\ref{a_8}), we can now write the Euclidean action $S_{0}$ of the
elastic model for the pure bubble crystal. We have 
\begin{equation}
S_{0}={\frac{1}{2T}}\sum_{\mathbf{k},\omega _{n}}\sum_{\alpha ,\beta
=x,y}u_{\alpha }\left( \mathbf{k},\omega _{n}\right) \,G_{\alpha \beta
}^{(0)-1}\left( \mathbf{k},\omega _{n}\,\right) u_{\beta }\left( -\mathbf{k}%
,-\omega _{n}\right) ,  \label{b1}
\end{equation}%
where the displacement fields obey the single Landau level dynamics \cite%
{kubo} 
\begin{equation}
\left[ u_{x}(\mathbf{R}),u_{y}(\mathbf{R}^{\prime })\right] =i\ell
^{2}\delta _{\mathbf{R},\mathbf{R}^{\prime }}.
\end{equation}%
From now on, we set $k_{B}=\hbar =1.$ In Eq. (\ref{b1}), $T$ is the
temperature, $\omega _{n}=2\pi n/T$ is the bosonic Matsubara frequency, and
the displacement Green's function $G_{\alpha \beta }^{(0)}\left( \mathbf{k}%
,i\omega _{n}\right) =\int_{0}^{1/T}d\tau e^{i\omega _{n}\tau
}\,\left\langle \mathcal{T}_{\tau }u_{\alpha }\left( \mathbf{k},\tau \right)
u_{\beta }\left( -\mathbf{k},0\right) \right\rangle _{S_{0}}$ is related to
the dynamical matrix by 
\begin{equation}
\widehat{G}^{(0)}\left( \mathbf{k},i\omega _{n}\right) =\frac{\ell ^{4}}{%
M^{2}\left( \omega _{n}^{2}+\omega _{\mathbf{k}}^{2}\right) }\left( 
\begin{array}{cc}
D_{yy}\left( \mathbf{k}\right) & \frac{M\omega _{n}}{\ell ^{2}}-D_{xy}\left( 
\mathbf{k}\right) \\ 
-\frac{M\omega _{n}}{\ell ^{2}}-D_{yx}\left( \mathbf{k}\right) & 
D_{xx}\left( \mathbf{k}\right)%
\end{array}%
\right) .  \label{b2}
\end{equation}%
Once the Green's function has been determined in the presence of disorder,
we can easily compute the conductivity. Since the electric current is
carried by the charge, the current density can be expressed as: 
\begin{equation}
\mathbf{j}\left( \mathbf{k},\tau \right) =i\frac{Me}{v_{c}}{\frac{d\mathbf{u}%
\left( \mathbf{k},\tau \right) }{d\tau }}.
\end{equation}%
The conductivity is then determined by the Kubo formula to be 
\begin{eqnarray}
\sigma _{\alpha \beta }\left( \omega \right) &=&-{\frac{1}{\omega \,S}}\,%
\left[ \int_{0}^{1/T}d\tau e^{i\omega _{n}\tau }\left\langle j_{\alpha
}\left( \mathbf{k}=0,\tau \right) j_{\beta }\left( \mathbf{k}=0,0\right)
\right\rangle \right] _{i\omega _{n}\rightarrow \omega +i0^{+}}  \label{b3}
\\
&=&-{\frac{M^{2}e^{2}}{v_{c}}}\,i\omega \,G_{\alpha \beta }^{\mathrm{ret}%
}\left( \mathbf{k}=0,\omega \right) ,  \notag
\end{eqnarray}%
where $S$ is the area of the 2DEG and $v_{c}$ is the unit cell of the bubble
crystal. Since $\widehat{D}\left( \mathbf{k}=0\right) =0$, it is easy to
check using Eqs. (\ref{b2}) and (\ref{b3}) that, in the pure limit and in
the strong magnetic field approximation, the electromagnetic response of the
system is purely transverse for all frequencies $\omega <<\omega _{c}$; 
\textit{i.e.}, we have in this limit\textit{\ }%
\begin{equation}
\widehat{\sigma }\left( \omega \right) =\left( 
\begin{array}{cc}
0 & \frac{nec}{B} \\ 
-\frac{nec}{B} & 0%
\end{array}%
\right) ,
\end{equation}%
where $n=M/\nu _{c}$ is the electronic density. When the mass term is not
neglected in Eq. (\ref{a_7}), we have instead%
\begin{equation}
\widehat{\sigma }\left( \omega \right) =\frac{ne^{2}/m^{\ast }}{\left(
\omega +i\delta \right) ^{2}-\omega _{c}^{2}}\left( 
\begin{array}{cc}
i\omega & -\omega _{c} \\ 
\omega _{c} & i\omega%
\end{array}%
\right) ,
\end{equation}%
and the absorption (which is proportional to $\Re \left[ \sigma _{xx}\left(
\omega \right) \right] $) is at the cyclotron frequency $\omega _{c}$ as it
should be from Kohn's theorem \cite{kohn}.

\subsection{Dynamical matrix in the GRPA}

In this work, we obtain the dynamical matrix from the GRPA. This procedures,
which uses the density response function $\chi _{\mathbf{K},\mathbf{K}%
^{\prime }}^{\left( \rho ,\rho \right) }\left( \mathbf{k},\omega +i\delta
\right) $ found microscopically to obtain the dynamical matrix for the
elastic theory, was developed to study pinned quantum Hall stripes, and is
described in detail in Ref. \onlinecite{meirong2}. Briely, we use the fact
that the density fluctuations are related to the displacement field via the
relation 
\begin{equation}
\delta n(\mathbf{k}+\mathbf{K},t)\approx -i\sqrt{N_{s}}h(\mathbf{k+K})\left( 
\mathbf{k+K}\right) \cdot \mathbf{u}(\mathbf{k}),
\end{equation}%
to obtain the relation 
\begin{equation}
\chi _{\mathbf{K},\mathbf{K}^{\prime }}^{\left( \rho ,\rho \right) }\left( 
\mathbf{k},\tau \right) =\frac{\nu }{M}\frac{h(\mathbf{k+K})h(\mathbf{k+K}%
^{\prime })}{F(\mathbf{k+K})F(\mathbf{k+K}^{\prime })}\sum_{\alpha ,\beta
}\left( k_{\alpha }+K_{\alpha }\right) G_{\alpha ,\beta }^{(0)}\left( 
\mathbf{k},\tau \right) \left( k_{\beta }+K_{\beta }^{\prime }\right)
\label{chiGF}
\end{equation}%
between the bare displacement Green's function $G_{\alpha \beta
}^{(0)}\left( \mathbf{k},\omega \right) $ and the density reponse function.
The density response function in the GRPA\ depends on the guiding center
densities $\left\langle \rho \left( \mathbf{K}\right) \right\rangle ^{\prime
}s$ computed in the Hartree-Fock approximation. The components of the matrix 
$\widehat{G}^{(0)}\left( \mathbf{k},\tau \right) $ (or of $\widehat{D}%
_{GRPA}\left( \mathbf{k}\right) $) are obtained from both the poles and the
weight of the density response function.

It is interesting to compare the dynamical matrix obtained in the GRPA from
that obtained with other approximations used in the literature. If the
filling factor $\nu $ is not too small, a rapidly convergent expression for
the \textit{classical} dynamical matrix of a lattice of rigid objects with
form factor $h\left( \mathbf{r}\right) $ is given by 
\begin{equation}
\widehat{D}_{\mathrm{classical}}\left( \mathbf{k}\right) =\frac{2\pi
n_{s}e^{2}}{\kappa }\sum_{\mathbf{K}}\left[ \frac{\left\vert h\left( \mathbf{%
k+K}\right) \right\vert ^{2}}{\left\vert \mathbf{k+K}\right\vert }\left( 
\mathbf{k+K}\right) \left( \mathbf{k+K}\right) -\frac{\left\vert h\left( 
\mathbf{K}\right) \right\vert ^{2}}{\left\vert \mathbf{K}\right\vert }%
\mathbf{KK}\right] ,  \label{a_10}
\end{equation}%
where $n_{s}=N_{s}/S$ is the density of crystal sites. This expression is
not suitable, however, for a lattice of point particles (p.p.). In this
case, we can use the expression of $\widehat{D}_{p.p}\left( \mathbf{k}%
\right) $ given by Bonsall and Maradudin\cite{maradudin} which is valid to
order $k^{2}$.

To compare the three expressions for $\widehat{D}\left( \mathbf{k}\right) $,
we consider the parameterization of the dynamical matrix in terms of the Lam%
\'{e} coefficients $\lambda $ and $\mu .$ The deformation energy of the
crystal can be expressed in terms of the dynamical matrix as $F=\frac{1}{2}%
\sum_{\alpha ,\beta }\sum_{\mathbf{k}}u_{\alpha }\left( \mathbf{k}\right)
D_{\alpha ,\beta }\left( \mathbf{k}\right) u_{\beta }\left( -\mathbf{k}%
\right) .$ On the other hand, the same energy is written\cite{LandauLifshitz}
in elastic theory (and for a triangular lattice) as $F=\frac{1}{2}\int d%
\mathbf{r}\left[ \lambda \left( e_{k,k}\right) ^{2}+2\mu e_{i,j}^{2}\right]
, $ where $e_{\alpha ,\beta }=\frac{1}{2}\left( \frac{\partial u_{\alpha
}\left( \mathbf{r}\right) }{\partial r_{\beta }}+\frac{\partial u_{\beta
}\left( \mathbf{r}\right) }{\partial r_{\alpha }}\right) $ is the strain
tensor. The coefficient $\mu $ is actually the shear modulus of the lattice
while the bulk modulus is given by $B=S\frac{\partial ^{2}F}{\partial S^{2}}%
=\lambda +\mu $. Comparing the two expressions for $F$, it is easy to see
that the dynamical matrix, in the long-wavelength limit, can be written in
terms of the Lam\'{e} coefficients as%
\begin{equation}
D_{\alpha ,\beta }\left( \mathbf{k}\right) =n_{s}^{-1}\left( \lambda +\mu
\right) k_{\alpha }k_{\beta }+n_{s}^{-1}\mu k^{2}\delta _{\alpha ,\beta }.
\label{a_11}
\end{equation}%
The long-range Coulomb force, absent in the elastic theory, is very
important for correctly capturing the low-energy physics. Its inclusion
leads to an additional term in $\lambda ,$ \textit{i.e. }to 
\begin{equation}
\lambda \rightarrow \left( \frac{2\pi n_{s}^{2}e^{2}}{\kappa }\right) \frac{%
M^{2}}{k}+\eta .
\end{equation}%
The long-wavelength dispersion relation of the magnetophonon mode is given
(to order $k^{2}$) by%
\begin{equation}
\omega =\frac{n_{s}^{-1}\ell ^{2}}{\hslash M}\sqrt{\left( \lambda +2\mu
\right) \mu }\,k^{2}.
\end{equation}%
It depends mostly on the shear modulus $\mu $ and on the long-range Coulomb
part of $\lambda .$ We will show later that the pinning behavior is also
mostly determined by these two quantities.

Using Eq. (\ref{a_11}), we can obtain $\mu _{GRPA}$ from our numerical data
for $\widehat{D}_{GRPA}\left( \mathbf{k}\right) $. The shear modulus $\mu _{%
\mathrm{classical}}$ can be obtained by making a small wavevector expansion
of Eq. (\ref{a_10}) taking into account the appropriate form factor $h\left( 
\mathbf{q}\right) $ that depends on $M$ and on the Landau level $N$. To
order $k^{3}$, we find that the Lam\'{e} coefficients can be written as%
\begin{eqnarray}
\lambda _{\mathrm{classical}} &=&\left( \frac{2\pi n_{s}^{2}e^{2}}{\kappa }%
\right) \left( \frac{M^{2}}{k}+\alpha _{0}+\alpha _{1}+2\alpha _{2}-\alpha
_{3}+\frac{\alpha _{4}}{3}+\beta _{0}k\right) , \\
\mu _{\mathrm{classical}} &=&\left( \frac{2\pi n_{s}^{2}e^{2}}{\kappa }%
\right) \left( \alpha _{3}+\frac{\alpha _{4}}{3}\right) .  \label{a_13b}
\end{eqnarray}%
The detailed forms of the coefficients $\alpha _{i}^{\prime }s$ and $\beta
_{i}^{\prime }s$ are given in Appendix A.

For a triangular lattice of point particles, we have instead\cite{maradudin}
(to order $k^{2}$) 
\begin{eqnarray}
\eta _{p.p} &=&-n_{s}\left( a+2b\right) , \\
\mu _{p.p} &=&n_{s}b,  \label{classicallimit}
\end{eqnarray}%
where $a=-1.225323\frac{M^{2}e^{2}}{\kappa \sqrt{v_{c}}},b=0.245065\frac{%
M^{2}e^{2}}{\kappa \sqrt{v_{c}}},$ with $v_{c}=\frac{\sqrt{3}}{2}a_{0}^{2}$
the unit cell area of the bubble crystal with $M$ electrons.

\begin{figure}[tbp]
\includegraphics[width=8cm]{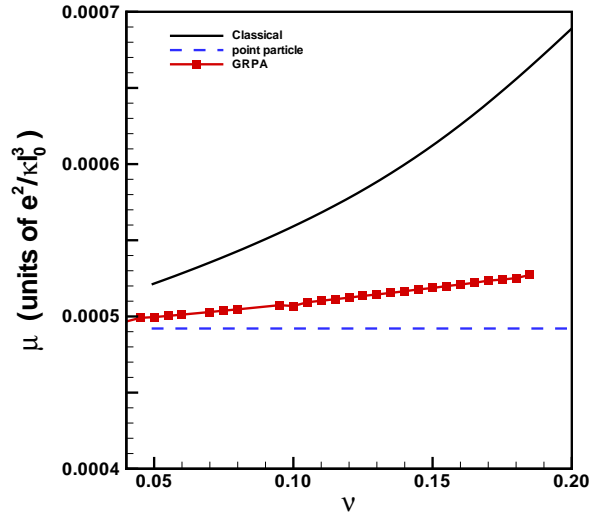}
\caption{Shear modulus computed in different approximations for Landau level 
$N=0$: for a classical lattice with form factor $h(\mathbf{G})$ (full line);
for a lattice of point particle (dashed line) and in the GRPA (line with
squares).}
\label{fig_shearmodulusn0}
\end{figure}

\begin{figure}[tbp]
\includegraphics[width=8cm]{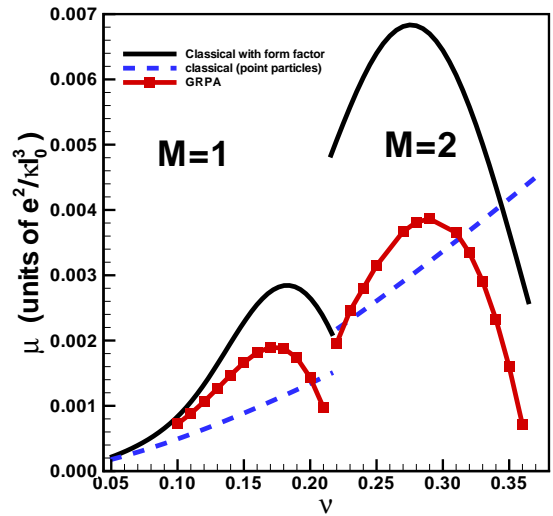}
\caption{Shear modulus computed in different approximations for Landau level 
$N=2$: for a classical lattice with form factor $h(\mathbf{G})$ (full line);
for a lattice of point particle (dashed line) and in the GRPA (line with
squares).}
\label{fig_shearmodulusn2}
\end{figure}

In Figs. \ref{fig_shearmodulusn0} and \ref{fig_shearmodulusn2} we compare
the three different approximations for the shear modulus. As can be seen
from Fig. \ref{fig_shearmodulusn0}, a point lattice is a good approximation
in Landau level $N=0$ (in comparison with the more exact GRPA\ result)
because the shear modulus does not vary much with filling factor. In $N=2$,
however, the form factor must be considered to account for the softening of
the lattice near the transition between two crystal phases. In this case,
the GRPA result is qualitatively similar to that of the classical
approximation although, quantitatively, the quantum corrections are
important especially when considering the jump in the shear modulus at the
transition. In the small $\nu $ limit, all three approximations give the
same result for $\mu $ as expected.

Note that in Fig. \ref{fig_shearmodulusn2}, the shear modulus decreases near 
$\nu _{2}$ where the Hartree-Fock calculation predicts a transition to
another bubble phase with $M=3$. In the transport experiments of Lilly et al.%
\cite{lilly}, there is no sign of such transition and, in the rest of this
paper, we will refer to the transition region as the region around filling
factor $\nu _{1}$. The range of filling factor where the $M=3$ is stable is
quite small in the HFA and, in a more accurate calculation, this phase is
probably unstable with respect to the stripe phase. Indeed, the $M=3$ bubble
state is absent in DMRG\cite{dmrg} calculations.

\section{Replicas and the Gaussian variational method}

\subsection{Disorder Coupling}

Now we take into account the disorder effect. We assume that the main source
of disorder comes from the interface roughness and that it can be
represented by a spatially uncorrelated Gaussian random potential $V(\mathbf{%
r})$. The disorder action reads 
\begin{equation}
S_{\mathrm{imp}}=\int d\mathbf{r}\,\int_{0}^{1/T}d\tau \,V\left( \mathbf{r}%
\,\right) n\left( \mathbf{r},\tau \right) ,  \label{b_4}
\end{equation}%
where $V\left( \mathbf{r}\right) $ has a Gaussian distribution function
which leads to the correlator 
\begin{equation}
\overline{V\left( \mathbf{r}_{1}\right) V\left( \mathbf{r}_{2}\right) }%
=V_{0}^{2}\,v_{c}\,\delta \left( \mathbf{r}_{1}-\mathbf{r}_{2}\right) .
\label{b_10}
\end{equation}%
Here the overline denotes an average over disorder configurations.

The electron density operator $n\left( \mathbf{r},\tau \right) $ in Eq.~({%
\ref{b_4}}) must be approximated more accurately than was needed in $G^{(0)}$
in order to capture the possibility of pinning by disorder. Following
Giamarchi and Le Doussal \cite{replica}, under the assumption of small $%
\nabla {\mathbf{u}}(\mathbf{r})$ (which is justified for weak disorder) we
write for the bubble crystals%
\begin{equation}
n\left( \mathbf{r},\tau \right) =n_{0}-n_{0}\mathbf{\nabla }\cdot \mathbf{u}%
\left( \mathbf{r},\tau \right) +n_{s}\sum_{\mathbf{K}\neq 0}h(\mathbf{K})e^{i%
\mathbf{K}\cdot \left[ \mathbf{r}-\mathbf{u}(\mathbf{r},\tau )\right] },
\end{equation}%
where $n_{0}=n_{s}h\left( 0\right) $ is the average electronic density.
Fourier transforming, we get 
\begin{equation}
n(\mathbf{q},\tau )\simeq \left[ N_{0}\delta _{\mathbf{q},0}-i\sqrt{N_{s}}M%
\mathbf{q}\cdot \mathbf{u}\left( \mathbf{k},\tau \right) +n_{s}\sum_{\mathbf{%
K}\neq 0}h(\mathbf{K})\int d\mathbf{r}\,e^{i\mathbf{K}\cdot \left[ \mathbf{r}%
-\mathbf{u}(\mathbf{r},\tau )\right] -i\mathbf{q}\cdot \mathbf{r}}\right] ,
\label{b_5}
\end{equation}%
where $\mathbf{q}=\mathbf{k}+\mathbf{K}$. Only the last term in Eq. (\ref%
{b_5}) captures the short wavelength oscillations in the charge density and
allows pinning by impurities\cite{replica} so that we will drop the first
two terms of this equation when using $n(\mathbf{q},\tau )$ in the disorder
action.

\subsection{Effective Form Factor for the Coupling with Disorder}

To simplify the coupling with disorder, it is usual to keep only the
summation over the first shell of reciprocal lattice vectors in Eq. (\ref%
{b_5}) (\textit{i.e.} $6$ vectors for a triangular lattice) as these give
the strongest restoring force for the displacement fields. Using Eqs. (\ref%
{a_1}) and (\ref{a_4}), one would then be tempted to use, for the form
factor $h(\mathbf{K})$ that enters Eq. (\ref{b_5}), the expression%
\begin{equation}
h(\mathbf{K})=\frac{M}{\nu }F_{N}(\mathbf{K})\left\langle \rho \left( 
\mathbf{K}\right) \right\rangle ,
\end{equation}%
where $\left\langle \rho \left( \mathbf{K}\right) \right\rangle $ is the
guiding-center density computed in the Hartree-Fock approximation. This
choice, however, underestimates the fluctuations in the potential as an
electron wavepacket or a bubble moves across the disorder potential. We can
capture this energy scale with an effective form factor, that we call $H(%
\mathbf{K}),$ in the following way. The density in the pure crystal is
exactly given by%
\begin{equation}
n\left( \mathbf{r}\right) =\frac{1}{S}\sum_{\mathbf{K}}n\left( \mathbf{K}%
\right) e^{i\mathbf{K}\cdot \mathbf{r}}.
\end{equation}%
The fluctuation in the disorder energy when a large patch of the crystal is
slid uniformly is then 
\begin{eqnarray}
\left\langle U^{2}\right\rangle &=&\int d\mathbf{r}\int d\mathbf{r}^{\prime
}\delta n\left( \mathbf{r}\right) \delta n\left( \mathbf{r}^{\prime }\right)
\left\langle \overline{V\left( \mathbf{r}\right) V\left( \mathbf{r}^{\prime
}\right) }\right\rangle , \\
&=&\frac{N_{s}V_{0}^{2}}{S^{2}}\sum_{\mathbf{K\neq 0}}\left\vert n\left( 
\mathbf{K}\right) \right\vert ^{2}.  \notag
\end{eqnarray}%
Using Eqs. (\ref{a_4}), (\ref{a_5}), and (\ref{a_1}) we find%
\begin{eqnarray}
\left\langle U^{2}\right\rangle &=&\frac{N_{s}V_{0}^{2}}{S^{2}}\sum_{\mathbf{%
K\neq 0}}\left\vert h\left( \mathbf{K}\right) \right\vert ^{2},  \label{b_6}
\\
&=&\frac{N_{s}V_{0}^{2}}{S^{2}}\frac{M^{2}}{\nu ^{2}}\sum_{\mathbf{K\neq 0}%
}\left\vert F_{N}\left( \mathbf{K}\right) \right\vert ^{2}\left\vert
\left\langle \rho \left( \mathbf{K}\right) \right\rangle \right\vert ^{2}, 
\notag
\end{eqnarray}%
where $\left\langle \rho \left( \mathbf{K}\right) \right\rangle $ is
computed in the Hartree-Fock approximation. On the other hand, if we
describe the fluctuation in the disorder energy by Eq. (\ref{b_5}), keeping
only one shell of reciprocal lattice vectors with modulus $K_{0},$ we get
for the same energy 
\begin{equation}
\left\langle U^{2}\right\rangle \approx \frac{N_{s}V_{0}^{2}}{S^{2}}%
6\left\vert H\left( K_{0}\right) \right\vert ^{2}.  \label{b_7}
\end{equation}%
We define our form factor, $H\left( K_{0}\right) ,$ by requiring both
expressions to be equal so that 
\begin{equation}
\left\vert H\left( K_{0}\right) \right\vert ^{2}=\frac{1}{6}\frac{M^{2}}{\nu
^{2}}\sum_{\mathbf{K\neq 0}}\left\vert F_{N}\left( \mathbf{K}\right)
\right\vert ^{2}\left\vert \left\langle \rho \left( \mathbf{K}\right)
\right\rangle \right\vert ^{2}.  \label{b_8}
\end{equation}%
In practice, we keep up to $50$ shells of reciprocal lattice vectors in the
summation on the right hand side of Eq. (\ref{b_8}).

From now on $\sum_{\mathbf{K}\neq 0}$ in the disorder action will mean a
summation over the first shell of reciprocal lattice vectors. This
simplification allows us to compute the Green's function in the presence of
disorder in a relatively straightforward manner while retaining the
essential physics of pinning so that our results are qualitatively correct.
The major effect of these approximations is to replace the soft cutoff in
wavevector that would enter through the form factor with a hard one in the
reciprocal lattice sum. With this approximation, the impurity action with
which we now work is 
\begin{subequations}
\begin{equation}
S_{\mathrm{imp}}=n_{s}\int d\mathbf{r}\,d\tau \,V\left( \mathbf{r}\right)
|H\left( K_{0}\right) |\,\sum_{\mathbf{K}\neq 0}e^{i\mathbf{K}\cdot \left[ 
\mathbf{r}-\mathbf{u}(\mathbf{r},\tau )\right] }.  \label{b_9}
\end{equation}

\subsection{Replicas and the GVM}

A detailed account of the replica and GVM can be found in Ref. %
\onlinecite{replica}. In previous work, we generalized this method to study
pinning of quantum Hall stripes near half filling in higher Landau levels~%
\cite{meirong1,meirong2}. Since the formalism can be applied with minor
changes to the pinning of the bubble crystals, we refer the reader to these
papers for details. In this section we only outline the procedure to obtain
the so-called saddle point equations (SPE's).

With the replica trick\cite{dotsenko}, one creates $n$ copies of the
original action, computes the replicated partition function $Z^{n}$, and
performs the disorder average on $Z^{n}$. Also by taking into account the
effective form factor as we discussed before, the resulting effective action
is 
\end{subequations}
\begin{eqnarray}
&&S_{\mathrm{eff}}=S_{0}^{(\mathrm{eff})}+S_{\mathrm{imp}}^{\mathrm{(eff)}},
\label{Seff} \\
&&S_{0}^{(\mathrm{eff})}={\frac{1}{2T}}\sum_{a=1}^{n}\sum_{\mathbf{k},\omega
_{n}}\sum_{\alpha ,\beta =x,y}u_{\alpha }^{a}\left( \mathbf{k},\omega
_{n}\right) \,G_{\alpha \beta }^{(0)-1}\left( \mathbf{k},\omega
_{n}\,\right) u_{\beta }^{a}\left( -\mathbf{k},-\omega _{n}\right) ,
\label{S0eff} \\
&&S_{\mathrm{imp}}^{\mathrm{(eff)}}\simeq -v_{\mathrm{imp}%
}\sum_{a,b=1}^{n}\,\int_{0}^{1/T}d\tau _{1}\int_{0}^{1/T}d\tau _{2}\,\sum_{%
\mathbf{r}}\sum_{\mathbf{K}\neq 0}\left\vert H\left( K_{0}\right)
\right\vert ^{2}  \label{Simpeff} \\
&&\times \cos \left[ \mathbf{K}\cdot \left[ \mathbf{u}^{a}(\mathbf{r},\tau
_{1})-\mathbf{u}^{b}(\mathbf{r},\tau _{2})\right] \right] ,  \notag
\end{eqnarray}%
where $v_{\mathrm{imp}}=V_{0}^{2}v_{c}^{2}n_{s}^{2}$, and $a,b$ are replica
indices that run from $1$ to $n$. In obtaining the last line of Eq.~(\ref%
{Simpeff}) we have neglected some rapidly oscillating terms.

In the pure limit, the action is diagonal in the replica indices. Disorder
averaging introduces a coupling among the replicas through the impurity
coupling $S_{\mathrm{imp}}^{\mathrm{(eff)}}$ in Eq.~(\ref{Simpeff}). This
coupling is non-Gaussian so we next apply the GVM to get a simplified
expression. The fundamental idea of the GVM is to replace $S_{\mathrm{eff}}$
with a variational action $S_{\mathrm{var}}$ that is \textit{quadratic}. We
write 
\begin{equation}
S_{\mathrm{var}}={\frac{1}{2T}}\sum_{\mathbf{k},\omega _{n}}u_{\alpha
}^{a}\left( \mathbf{k},\omega _{n}\right) \,\left( G^{-1}\right) _{\alpha
\beta }^{ab}\left( \mathbf{k},\omega _{n}\right) \,u_{\beta }^{b}\left( -%
\mathbf{k},-\omega _{n}\right) ,  \label{varaction}
\end{equation}%
with the coefficients $\left( G^{-1}\right) _{\alpha \beta }^{ab}\left( 
\mathbf{k},\omega _{n}\right) $ chosen to best match the original problem.
This is accomplished by minimizing a free energy \cite{replica} $F_{\mathrm{%
var}}=F_{0}+T\left[ \left\langle S\right\rangle _{S_{\mathrm{var}%
}}-\left\langle S_{\mathrm{var}}\right\rangle _{S_{\mathrm{var}}}\right] $
where $F_{0}$ is the free energy associated with the action $S_{\mathrm{var}%
} $, and $\left\langle \cdots \right\rangle _{S_{\mathrm{var}}}$ indicates a
functional integral over displacements, with $S_{\mathrm{var}}$ as a
weighting function. In Eq.~(\ref{varaction}), $G_{\alpha \beta }^{ab}\left( 
\mathbf{k},\omega _{n}\right) $ is the displacement Green's function, 
\begin{equation}
G_{\alpha \beta }^{ab}\left( \mathbf{k},\omega _{n}\right)
=\int_{0}^{1/T}d\tau \left\langle T_{\tau }u_{\alpha }^{a}\left( \mathbf{k}%
,\tau \right) u_{\beta }^{b}\left( -\mathbf{k},0\right) \right\rangle _{S_{%
\mathrm{var}}}.
\end{equation}%
It is convenient to write it in terms of the bare Green's function via the
equation%
\begin{equation}
\left( G^{-1}\right) _{\alpha \beta }^{ab}\left( \mathbf{k},\omega
_{n}\right) =G_{\alpha \beta }^{(0)-1}\left( \mathbf{k},\omega _{n}\right)
\,\delta _{ab}-\zeta _{\alpha \beta }^{ab}\left( \omega _{n}\right) ,
\end{equation}%
where $\zeta _{\alpha \beta }^{ab}\left( \omega _{n}\right) $ is the element
of the variational self-energy matrix $\hat{\zeta}$. Note that there is no $%
\mathbf{k}$ dependence in $\hat{\zeta}$ because we have chosen our impurity
action to be local in space; this will become clear when we find the SPE's
below. Note also the obvious symmetries $G^{ab}=G^{ba}$ and $\zeta
^{ab}=\zeta ^{ba}$.

\subsection{Saddle Point Equations (SPE's)}

Finding the extremum of the free energy leads to the saddle point equations
(SPE's) 
\begin{eqnarray}
&&\zeta _{\alpha \beta }^{aa}(\omega _{n})=4v_{\mathrm{imp}%
}\int_{0}^{1/T}d\tau \,\left\{ \left( 1-\cos \omega _{n}\tau \right)
\,V_{\alpha \beta }^{\prime }\left[ B^{aa}(\tau )\right] +\sum_{b\neq
a}V_{\alpha \beta }^{\prime }\left[ B^{ab}(\tau )\right] \right\} ,
\label{zetaaa} \\
&&\zeta _{\alpha \beta }^{a(b\neq a)}(\omega _{n})=-4v_{\mathrm{imp}%
}\int_{0}^{1/T}d\tau \,\cos \omega _{n}\tau \,V_{\alpha \beta }^{\prime }%
\left[ B^{ab}(\tau )\right] ,  \label{zetaab}
\end{eqnarray}%
where 
\begin{equation}
V_{\alpha \beta }^{\prime }\left[ B^{ab}(\tau )\right] =\sum_{\mathbf{K}\neq
0}\left\vert H\left( K_{0}\right) \right\vert ^{2}K_{\alpha }K_{\beta
}\,\exp \left[ -{\frac{1}{2}}\sum_{\mu \nu =x,y}K_{\mu }K_{\nu }B_{\mu \nu
}^{ab}(\tau )\right] ,  \label{s_15}
\end{equation}%
and%
\begin{eqnarray}
B_{\alpha \beta }^{ab}(\tau ) &=&\left\langle T_{\tau }[u_{\alpha }^{a}(%
\mathbf{r},\tau )-u_{\beta }^{b}(\mathbf{r},0)]^{2}\right\rangle _{S_{%
\mathrm{var}}}  \label{Bab} \\
&=&T\frac{1}{N_{s}}\sum_{\mathbf{k},\omega _{n}}\left[ G_{\alpha \beta
}^{aa}(\mathbf{k},\omega _{n})+G_{\alpha \beta }^{bb}(\mathbf{k},\omega
_{n})-2\cos (\omega _{n}\tau )G_{\alpha \beta }^{ab}(\mathbf{k},\omega _{n})%
\right] .  \notag
\end{eqnarray}

It is apparent at this point that the self-energy has no $\mathbf{k}$
dependence. Moreover, if we assume that reflection symmetry for the bubble
crystals are not spontaneously broken after disorder averaging, then the
solutions of interest to Eqs.~(\ref{zetaaa}) and (\ref{zetaab}) will satisfy 
$\zeta _{xy}^{ab}=0$ and $\zeta _{xx}^{ab}=\zeta _{yy}^{ab}$. Using the
symmetry [see Eq. (\ref{a_11})] of the dynamical matrix $D_{xx}\left(
k_{x},k_{y}\right) =D_{yy}\left( k_{y},k_{x}\right) $ for a triangular
lattice, we can also show that $\zeta _{xx}^{ab}=\zeta _{yy}^{ab}$ is a
solution of the SPE's. We need thus solve for one self-energy component only.

At this point, we take the limit $n\rightarrow 0$ in the replica index. In
taking this limit, one \textit{assumes} that the self-energy and Green's
function matrices can be written in a \textquotedblleft hierarchical
form\textquotedblright\ \cite{dotsenko}. A consequence of this is that the
replica index becomes a continuous variable $u\in \left[ 0,1\right] $ (The
variable $u$ must not be confused with the displacement field here.) For
example, the self-energy matrix is now characterized by a diagonal
component, $\tilde{\zeta}_{\alpha }$, and an off-diagonal component, $\zeta
_{\alpha }(u),$ such that 
\begin{eqnarray}
&&\zeta _{\alpha \alpha }^{aa}\rightarrow \tilde{\zeta}_{\alpha }, \\
&&\zeta _{\alpha \alpha }^{ab(\neq a)}\rightarrow \zeta _{\alpha }(u),\;\;\;%
\mathrm{for}\;\;0\leq u\leq 1.
\end{eqnarray}%
Similarly, $G_{\alpha \beta }^{aa}\rightarrow \widetilde{G}_{\alpha \beta }$%
, $G_{\alpha \beta }^{ab(\neq a)}\rightarrow G_{\alpha \beta }(u)$ ($0\leq
u\leq 1$). Since the disorder potential $V(\mathbf{r})$ is time independent,
a further simplification one finds is that the off-diagonal replica
components $\zeta _{\alpha \alpha }^{ab(\neq a)}$ and $G_{\alpha \beta
}^{ab(\neq a)}$ are $\tau $ independent\cite{replica} so that $\widehat{G}(%
\mathbf{k},\omega _{n},u)$ and $\hat{\zeta}(\mathbf{k},\omega _{n},u)$ are
different from zero only for $\omega _{n}=0.$ We have 
\begin{eqnarray}
&&\widehat{G}(\mathbf{k},\omega _{n},u)=\widehat{G}(\mathbf{k},u)\,\delta
_{\omega _{n},0},  \label{nown1} \\
&&\hat{\zeta}(\omega _{n},u)=\hat{\zeta}(u)\,\delta _{\omega _{n},0}.
\label{nown2}
\end{eqnarray}%
The SPE's (\ref{zetaaa}) and (\ref{zetaab}) for the \textit{diagonal
component} $\widetilde{\zeta }_{xx}=\widetilde{\zeta }_{yy}\equiv \widetilde{%
\zeta }$ may now be written as 
\begin{eqnarray}
&&\tilde{\zeta}(\omega _{n})=\int_{0}^{1}du\,\zeta (u)+4v_{\mathrm{imp}%
}\int_{0}^{1/T}d\tau \left( 1-\cos \left( \omega _{n}\tau \right) \right)
\,V_{xx}^{\prime }\left[ \widetilde{B}(\tau )\right] ,  \label{zetatilde} \\
&&\zeta (u)=-{\frac{4v_{\mathrm{imp}}}{T}}V_{xx}^{\prime }\left[ B(u)\right]
,  \label{zetau}
\end{eqnarray}%
where, from Eq.~(\ref{Bab}), 
\begin{eqnarray}
&&\widetilde{B}_{xx}(\tau )=2T\sum_{\mathbf{k},\omega _{n}}\left( 1-\cos
(\omega _{n}\tau )\right) \widetilde{G}_{xx}(\mathbf{k},\omega _{n}),
\label{Btilde} \\
&&B_{xx}(u)=2T\sum_{\mathbf{k}}\left\{ \left[ \sum_{\omega _{n}}\widetilde{G}%
_{xx}(\mathbf{k},\omega _{n})\right] -G_{xx}(\mathbf{k},u)\right\} .
\end{eqnarray}%
Note that Eq.~(\ref{zetatilde}) also gives us $\tilde{\zeta}(\omega
_{n}=0)=\int_{0}^{1}du\zeta (u).$

To compute the dynamical conductivities in Eq.~(\ref{b3}), we first need to
solve the SPE's for the finite-frequency self energy $\tilde{\zeta}(\omega
_{n})$. If we take the analytical continuation of Eq. (\ref{zetatilde}), we
get 
\begin{eqnarray}
&&\tilde{\zeta}^{\mathrm{ret}}(\omega )=e_{0}-24v_{\mathrm{imp}}\left\vert
H\left( K_{0}\right) \right\vert ^{2}K_{0}^{2}\int_{0}^{\infty
}dt(e^{i\omega t}-1)  \notag \\
&&\;\;\;\;\;\;\;\;\times \mathrm{Im}\exp \left\{ -{\frac{K_{0}^{2}}{\pi }}%
\int_{0}^{\infty }df\mathrm{Im}\left[ \sum_{\mathbf{k}}\tilde{G}_{xx}^{%
\mathrm{ret}}(\mathbf{k},\omega )\right] (1-e^{-ift})\right\}
\label{complex}
\end{eqnarray}%
where $e_{0}=\tilde{\zeta}^{\mathrm{ret}}(0^{+})$ is a constant which will
be determined below. Eqs.~(\ref{nown1}) and (\ref{nown2}) indicate that for
non-zero frequency $\omega _{n},$ 
\begin{equation}
\widetilde{G}_{xx}(\mathbf{k},\omega _{n}\neq 0)=\left[ \widehat{G}^{(0)-1}(%
\mathbf{k},\omega _{n})-\hat{\tilde{\zeta}}(\omega _{n})\right] _{xx}^{-1}.
\label{tildeG}
\end{equation}%
Thus, once we have obtained the self-energy, we can easily compute the
finite-frequency conductivities in Eq.~(\ref{b3}) by analytically continuing
Eq.~(\ref{tildeG}) to real frequency. The retarded Green's function for the
displacement field is given by 
\begin{equation}
\widetilde{G}_{xx}^{\mathrm{ret}}(\mathbf{k},\omega \neq 0)=\left[ \widehat{G%
}_{\mathrm{ret}}^{(0)-1}(\mathbf{k},\omega )-\hat{\tilde{\zeta}}^{\mathrm{ret%
}}(\omega )\right] _{xx}^{-1},
\end{equation}%
and so%
\begin{equation}
G\left( \mathbf{k},\omega \right) =\frac{\left( 
\begin{array}{cc}
D_{yy}\left( \mathbf{k}\right) -\tilde{\zeta}^{\mathrm{ret}}\left( \omega
\right) & \frac{-i\omega M}{\ell ^{2}}-D_{xy}\left( \mathbf{k}\right) \\ 
\frac{i\omega M}{\ell ^{2}}-D_{yx}\left( \mathbf{k}\right) & D_{xx}\left( 
\mathbf{k}\right) -\tilde{\zeta}^{\mathrm{ret}}\left( \omega \right)%
\end{array}%
\right) }{\det \left[ D\right] -\tilde{\zeta}^{\mathrm{ret}}\left( \omega
\right) \mathrm{tr}\left[ D\right] +\left( \tilde{\zeta}^{\mathrm{ret}%
}\left( \omega _{n}\right) \right) ^{2}-M^{2}\omega ^{2}/\ell ^{4}}.
\label{green2}
\end{equation}%
Inserting this result into Eq. (\ref{b3}), we obtain our final expression
for the longitudinal conductivity in the presence of disorder, 
\begin{equation}
\sigma _{xx}(\omega )=\sigma _{yy}(\omega )={\frac{M^{2}e^{2}}{v_{c}}}{\frac{%
i\omega \tilde{\zeta}^{\mathrm{ret}}(\omega )}{\left[ \tilde{\zeta}^{\mathrm{%
ret}}(\omega )\right] ^{2}-M^{2}\omega ^{2}/\ell ^{4}}.}
\label{longitudinalconductivity}
\end{equation}

\subsection{Semiclassical approximation}

Equation (\ref{complex}) for the self-energy is a very complex
self-consistent equation. In studying the stripe phase\cite%
{meirong1,meirong2}, where a depinning transition may take place and
relative fluctuations in the displacement field can become very large, we
found out that it was necessary to solve this equation exactly. In the case
of the bubble crystals, the transition between the bubble phases are first
order and we expect that the fluctuations in the relative displacement of
the bubbles are small enough so that we can use a semiclassical version of
Eq. (\ref{complex}). The semiclassical expression for the self-energy\cite%
{replica} is obtained by expanding the exponential in Eq. (\ref{s_15}) and
keeping only the leading term. This leads to 
\begin{equation}
\tilde{\zeta}^{\mathrm{ret}}(\omega )=e_{0}+\frac{\Delta }{N_{s}}\sum_{%
\mathbf{k}}\left[ \tilde{G}_{xx}^{\mathrm{ret}}(\mathbf{k},\omega )-\tilde{G}%
_{xx}^{\mathrm{ret}}(\mathbf{k},\omega =0^{+})\right] ,
\label{semiclassical}
\end{equation}%
where%
\begin{equation}
\Delta =12v_{\mathrm{imp}}\left\vert H\left( K_{0}\right) \right\vert
^{2}K_{0}^{4}.  \label{delta}
\end{equation}

The semiclassical approximation is a powerful simplification when it is
valid. In particular one sees that Eq.~(\ref{semiclassical}) is local in the
frequency $\omega $ so that $\tilde{\zeta}^{\mathrm{ret}}(\omega )$ can be
determined one frequency at a time. As we will see later, $e_{0}\neq 0$ is
an energy offset that, in a pinned state, opens a gap in the phonon
spectrum.\qquad

\subsection{A Constraint for $e_{0}$}

The SPE (\ref{semiclassical}) becomes a set of two coupled equations for the
real and imaginary parts of the self-energy if $e_{0}$ is known. Formally, $%
e_{0}$ needs to be determined self-consistently by solving Eq.~(\ref%
{semiclassical}) together with those for $\zeta (u)$. The SPE's have a
family of solutions parameterized by $u_{c}$ where $u_{c}$ is, in principle,
determined by minimizing the free energy. In the case of spatial dimension $%
d>2$, $u_{c}$ determined in this way leads to $\Re \left[ \sigma _{\alpha
\alpha }(\omega )\right] \sim \omega ^{2}$ at small $\omega $. This behavior
is consistent with arguments by Mott as well as with some exact solutions 
\cite{exactsolution} (up to a logarithmic correction). For $d\leq 2$,
however, this way of finding $u_{c}$ can yield an unphysical result in
which, in the pinned state, the conductivity shows a true gap \textit{i.e.} $%
\Re \left[ \sigma \left( \omega \right) \right] $ vanishes below some finite
frequency. For $d\leq 2$, what can instead be done is to \textit{impose} the
condition $\Re \left[ \sigma _{\alpha \alpha }(\omega )\right] \sim \omega
^{2}$ at small $\omega $. This common procedure, although not fully
understood, leads to physically reasonable results and we will adopt it in
the computations that follow. (The reader should keep in mind that it leads
to a broader response in the WC regime than is found by other methods. We
discuss this issue below.) From Eq.~(\ref{longitudinalconductivity}), this
condition is equivalent to the condition $\Im \left[ \widetilde{\zeta }^{%
\mathrm{ret}}(\omega )\right] \sim \omega $ at small $\omega $ and this
guarantees that the magnetophonon mode density of states vanishes at zero
frequency as it should for a pinned system.

To determine $e_{0}$, we thus write, for small $\omega $,%
\begin{equation}
\tilde{\zeta}_{ret}\left( \omega \right) \approx e_{0}+i\beta \omega .
\end{equation}%
With this constraint in Eq. (\ref{semiclassical}), using Eq. (\ref{green2})
and the symmetry relation $\tilde{\zeta}_{xx}^{\mathrm{ret}}(\omega )=\tilde{%
\zeta}_{yy}^{\mathrm{ret}}(\omega )$, we can easily show that the condition
for nonvanishing $\beta $ leads to%
\begin{equation}
1=-\Delta \sum_{\mathbf{k}}\left[ \frac{\det \left[ D\right] +e_{0}\mathrm{tr%
}\left[ D\right] -e_{0}^{2}-\frac{1}{2}\left( \mathrm{tr}\left[ D\right]
\right) ^{2}}{\left( \det \left[ D\right] -e_{0}\mathrm{tr}\left[ D\right]
+e_{0}^{2}\right) ^{2}}\right] .  \label{constraint}
\end{equation}%
Eq.~(\ref{constraint}) is the constraint which we solve numerically to get $%
e_{0}$.

\section{Numerical results}

\begin{figure}[tbp]
\includegraphics[width=8cm]{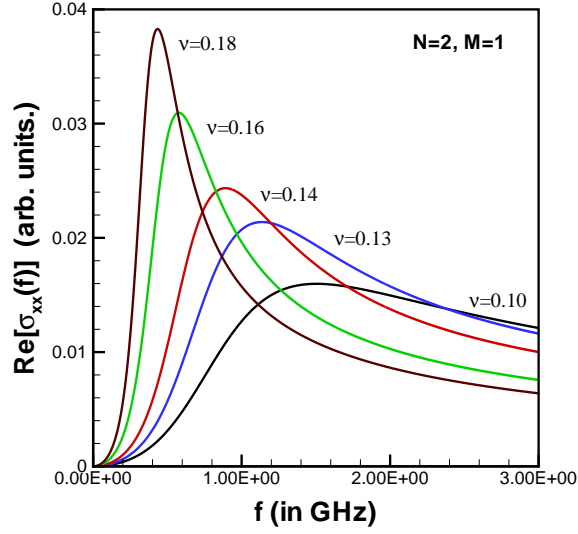}
\caption{Dynamical conductivity for the Wigner crystal in Landau level $N=2$%
. }
\label{sigman2m1}
\end{figure}

\begin{figure}[tbp]
\includegraphics[width=8cm]{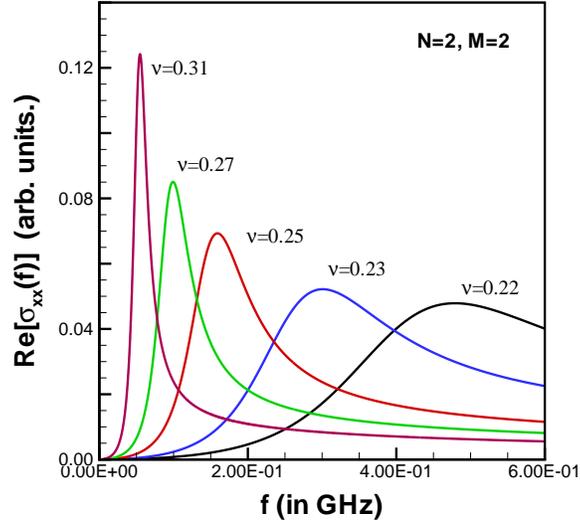}
\caption{Dynamical conductivity for the $M=2$ bubble crystal in Landau level 
$N=2.$ }
\label{sigman2m2p}
\end{figure}

We now discuss our numerical results for the behavior of the dynamical
conductivity $\sigma _{xx}\left( \omega \right) $ with filling factor in
Landau level $N=2$. Figures \ref{sigman2m1} and \ref{sigman2m2p} show the
real part of the dynamical conductivity $\Re \left[ \sigma _{xx}\left(
\omega \right) \right] $ for different values of the filling factor $\nu $
in the $M=1$ and $M=2$ bubble phases. At the transition between two bubble
phases, the density is such that the cyclotron radius $R_{c}=\sqrt{2N+1}\ell 
$ of adjacent bubbles touches one another. Away from the transition region,
however, the size of the bubbles is smaller than the lattice spacing. For
all filling factors here we assume the bubbles are much bigger than the
correlation length of the disorder (which we take as zero).

\begin{figure}[tbp]
\includegraphics[width=8cm]{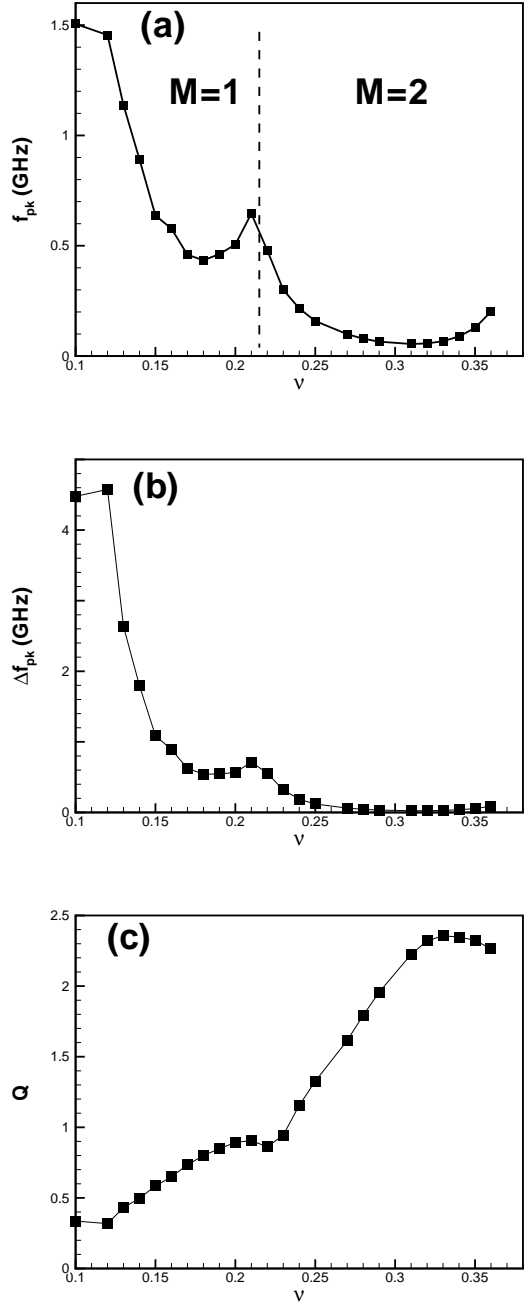}
\caption{Numerical results for the bubble phases in Landau level $N=2$ for
(a) the pinning peak; (b) the width of the pinning peak; (c) the quality
factor.}
\label{combinedn2}
\end{figure}

We plot in Fig. \ref{combinedn2} (a)\ the peak (or pinning) frequency $%
f_{pk}=2\pi \omega _{pk}$, (b) the full width at half maximum of the
resonance, $\Delta f_{pk}$, and (c) the quality factor $Q=f_{pk}/\Delta
f_{pk}$. Although we do not expect our theory to be quantitatively accurate,
we choose the disorder level in our calculation so that our pinning
frequency at $\nu =0.1$ is equal to that measured experimentally\cite{lewis2}
at the same filling factor; \textit{i.e.}, $f_{pk}\approx 1.5$ GHz with an
electronic density of $n=3.0\times 10^{11}$ cm$^{-2}$. We find that our
results are in a very good qualitative agreement with those of Lewis et al.%
\cite{lewis2} with the exception of the region between $\nu _{a}=0.16$ to $%
\nu _{b}=0.28$ where the two phases are assumed to coexist.

The variation of the conductivity with $\nu $ is the same for the $M=1$ and $%
M=2$ phases. In both cases the conductivity curve has a non-Lorentzian
lineshape with a peak frequency and width that decrease with increasing
filling factor $\nu $ (or density) and a peak height that increases with
increasing filling factor. This behavior is qualitatively consistent with
that of a weak-pinning model. Quantitatively, however, our pinning frequency
decreases faster (we find $f_{pk}\sim 1/\nu ^{3}$, for $M=1$) than what is
seen experimentally where the dependence $f_{pk}\sim 1/\nu ^{3/2}$ is found.
The width of the pinning peak and the quality factor that we get, however,
are typical of what are measured experimentally.

If we compare the behavior of the pinning frequency in Fig. \ref{combinedn2}
with that of the shear modulus calculated in Fig. \ref{fig_shearmodulusn2},
we see that the pinning frequency behavior is opposite to that of the shear
modulus within the whole range of filling factor considered. A more rigid
crystal is less sensitive to the disorder potential and has, as a
consequence, a smaller pinning frequency. This is precisely what is expected
from collective pinning theory \cite{larkin}.

The increase in the pinning frequency that we get after $\nu \approx 0.18$
in the $M=1$ phase is not seen experimentally. Instead, two pinning
frequencies separated by a gap of the order of $0.2$ GHz are detected in the
region between the dashed lines in Fig. \ref{fig_energien2}. These two
frequencies are found to be nearly frequency independent in this range (%
\textit{i.e.} from $\nu _{a}=0.16$ to $\nu _{b}=0.28$) where the two crystal
phases probably coexist. Interestingly, the difference in pinning
frequencies at the endpoints of this interval, $f_{pk}\left( \nu _{a}\right)
-f_{pk}\left( \nu _{b}\right) ,$ is $0.42$ GHz in our calculation. We remark
that the choice of the form factor $H\left( \mathbf{K}_{0}\right) $ [see Eq.
(\ref{b_8})] that enters in the coupling with the disorder [Eq. (\ref%
{Simpeff})] is important in getting the right order of magnitude for the
drop in the pinning frequency. We find, for example, that replacing the $%
\sum_{\mathbf{K}\neq 0}^{\prime }\left\vert H\left( K_{0}\right) \right\vert
^{2}\left( \ldots \right) $ in Eq. (\ref{Simpeff}) (the sum is here on the
first shell of reciprocal lattice vectors only) by the more na\"{\i}ve form $%
\sum_{\mathbf{K}\neq 0}\left\vert h(\mathbf{K})\right\vert ^{2}\left( \ldots
\right) $ (where the sum is over all reciprocal lattice vectors)\ and using
a cutoff in $\mathbf{K}$ in this last expression overestimate the sudden
reduction in the pinning frequency seen experimentally\cite{lewis2}. The
formalism we use needs to be generalized in order for us to properly compute
the dynamical conductivity in the coexistence region.

In Landau level $N=0,$ our numerical calculations give results similar to
those of Chitra and coworkers.\cite{chitra}. For uncorrelated disorder, the
pinning frequency increases continuously with decreasing filling factor (at
fixed density). For a disorder level ($v_{imp}$ in our calculation)
corresponding to the typical peak frequencies observed experimentally\cite%
{ye}, we get a quality factor $Q$ of the order of $1.$We have found that if
one solves the GVM equations for \textit{very} small disorder values in
Landau level $N=0$, a sharp resonance emerges with $Q\sim \Delta ^{-1/4}$.
However, this behavior appears only at disorder strengths yielding resonance
frequencies far below those observed in experiment.

We note that experiments on higher density \textit{hole} systems \cite%
{ccli,mellor} have revealed dramatically narrower resonances, a phenomenon
that our present calculations do not seem to reproduce. We discuss this
issue in the next section.

\section{Long Range Coulomb Interaction and Width of the Pinning Peak}

One issue regarding the lineshape of the dynamical conductivity near the
pinning frequency involves the observation in hole systems, in the Wigner
crystal regime, of pinning peaks with quality factor $Q$ of order $5$ and
even higher \cite{ccli,mellor}.

The possibility of such dramatic narrowing was suggested by one of us \cite%
{fertig99} as being a result of the long-range Coulomb interaction, in
conjunction with the effect of the magnetic field. In Ref. %
\onlinecite{fertig99}, a simulation of an electron system in a strong
magnetic field with a simplified disorder model yielded a \textit{remarkably}
sharp resonance, with width so narrow it could not be determined
numerically. That this effect was related to the long-range nature of the
Coulomb interaction was demonstrated by running the same simulation with a
screened interaction, which yielded a more typical pinning peak with $Q\sim
1 $.

How such a narrow resonance might be attained can be understood in a
self-consistent Born approximation (SCBA). Writing the self-energy as $\zeta
(\omega )=z_{1}+iz_{2}$, the imaginary part $z_{2}(\omega )$ at the
resonance frequency sets the width of the pinning peak. The SCBA can be
written in the form 
\begin{equation}
\zeta (\omega )=\Delta \int d^{2}q~G_{xx}^{ret}(\mathbf{q},\omega ),
\label{scba}
\end{equation}%
where $\Delta $ is the scale of the disorder interaction defined in Eq. (\ref%
{delta}). The similarity of this equation with the semiclassical
approximation in the GVM [Eq. (\ref{semiclassical})] is apparent, and indeed
if one takes the imaginary part of Eq. (\ref{scba}) it is identical to the
imaginary part of Eq. (\ref{semiclassical}). This equation can be written
explicitly in the form 
\begin{equation}
1=\Delta \int d^{2}q{\frac{{{\frac{1}{2}}}\left( {tr}\left[ D\right] \right)
^{2}{-\det }\left[ D\right] {+\omega ^{2}-z_{1}{tr}\left[ D\right]
+z_{1}^{2}+z_{2}^{2}}}{\left[ {\det }\left[ {D}\right] {%
+z_{1}^{2}-z_{2}^{2}-\omega ^{2}-z_{1}{tr}\left[ D\right] }\right] {%
^{2}+z_{2}^{2}}\left[ {{tr}\left[ D\right] -2z_{1}}\right] {^{2}}}}.
\label{imscba}
\end{equation}
A solution to Eq. (\ref{imscba}) appears at first challenging for small $%
\Delta $ because the right hand side appears to vanish as $\Delta
\rightarrow 0$ whereas the left hand side remains at one. A solution is
always possible for any given value of $z_{1}$ because the integral is
logarithmically divergent as $z_{2}\rightarrow 0$. This can be seen by
noting that at small $q$, the trace and determinant respectively have the
forms ${{tr}\left[ D\left( \mathbf{q}\right) \right] }=Yq$ and ${\det }\left[
D\left( \mathbf{q}\right) \right] =\alpha q^{3}$. (Note for short range
interactions the powers of $q$ in each of these expressions increase by $1$%
.) If we choose $\omega =-z_{1}(\omega )$, \textit{i.e.,} the resonance
frequency, the small frequency limit of the integrand gives 
\begin{equation}
1\approx 4\pi \Delta \int_{0}^{q_{c}}dq{\frac{q}{{Y^{2}q^{2}+4z_{2}^{2}}}},
\label{smqsca}
\end{equation}%
with $q_{c}=\sqrt{z_{1}Y/\alpha }$. We then arrive at a solution of the form 
\begin{equation}
z_{2}\approx {\frac{{Yq_{c}}}{2}}e^{-Y^{2}/4\pi \Delta }.  \label{z2exp}
\end{equation}%
A very similar result was obtained in Ref. \onlinecite{fertig99} using a
different method. The exponential behavior in Eq. (\ref{z2exp}) allows for a
very narrow resonance, as was found in the simulation. As discussed in Ref. %
\onlinecite{fertig99}, the result may be understood as being due to a
suppression of the collective mode density of states at low frequencies,
arising from the effectively very large stiffness that is a result of the
Coulomb interaction.

Given the similarity between Eqs. (\ref{imscba}) and (\ref{semiclassical}),
it is surprising that the GVM approach does not obtain this narrow
resonance. The reason for this difference can be understood by the way in
which the GVM is \textit{different} from the SCBA, specifically by how $e_{0}
$ is determined. By definition, $e_{0}=z_{1}(\omega \rightarrow 0)$, which
is obtained in the SCBA by self-consistently solving the real part of Eq. (%
\ref{scba}). A major limitation of this approach is that it does not
correctly capture under most circumstances the competition between
elasticity and pinning energy that sets the pinning frequency \cite%
{fukulee,larkin}. In the GVM, $e_{0}$ is obtained by solving Eq. (\ref%
{constraint}), which followed from requiring $\Re \left[ \sigma _{\alpha
\alpha }(\omega )\right] \sim \omega ^{2}$ at small $\omega $. There is a
very close connection between Eqs. (\ref{constraint}) and (\ref{smqsca}):
the integrands are \textit{identical} for $q/q_{c}>>1$. Because $q_{c}$
becomes very small for small disorder, by noting that $z_{1}(\omega )$ is
not very different than $e_{0}$ near the resonance we can see the choice of $%
e_{0}$ \textit{guarantees} that Eq. (\ref{imscba}) will very nearly be
satisfied from the portion of the integral where $|\mathbf{q}|\geq q_{c}.$%
The small differences in the two equations can then be accomodated with a
value of $z_{2}$ not very different than $z_{1}.$This leads to the more
typical result for a pinned system, $Q\sim 1$.

We conclude this section with a few comments. First, it should be noted that
the very sharp resonances predicted by the SCBA and related work \cite%
{fertig99,narrowpeak} are \textit{not} what is observed experimentally.
While the hole experiments yield peaks with a surprisingly large $Q$, they
are many orders of magnitude smaller than what is predicted by Eq. (\ref%
{z2exp}). If such a very narrow resonance is the correct expectation for the
pinned Coulomb system, then there is presumably some dissipation mechanism
that has not been identified. One cannot help but notice, however, that the
results of replicas+GVM method are far more similar to the experimental
results than the SCBA by itself gives. And while the simulation reported in
Ref. \onlinecite{fertig99} suggests a very narrow resonance, it is possible
that the simplified disorder model used there may give a different result
than a generic disorder model (although previous studies have suggested this
should not be the case~\cite{normand}.)

Secondly, it should be emphasized that the choice of $e_0$ in Eq. (\ref%
{constraint}) really is very special. If $e_0$ is allowed to vary even by a
very small amount from the solution of this equation, we have verified that
the semiclassical approximation to the SPE's [Eq. (\ref{semiclassical})]
results in a very sharp resonance as expected from the SCBA. This
demonstrates that the important difference between the method used in this
paper and the SCBA is not really in the use of replicas or the GVM, but
rather the introduction of the extra constraint. Indeed, it was suggested by
Fogler and Huse \cite{narrowpeak} that if one can instead impose a condition 
$\Re \left[ \sigma _{\alpha \alpha }(\omega )\right] \sim \omega^x$ with a
large value of $x$, then the $Q$ of the resulting resonance is again
extremely large -- smaller than what is expected from the SCBA, but still
orders of magnitude larger than what is seen in experiment.

A third important point is that, although these considerations leave us with
questions about the $Q$ of the resonance, it seems likely the resonance
frequency itself is well-represented by the calculations described in this
work. This is because the pinning resonance is determined by collective
pinning, which balances the lowest two energy scales -- elastic and disorder
interactions -- to determine the restoring force for motion of the system as
a whole. (Again, we emphasize that even a very small adjustment of $e_0$
results in a narrow resonance. The resonance frequency does not need to
change noticeably to dramatically increase $Q$.) A resonance frequency
determined by collective pinning is consistent with what was found in the
simulation of Ref. \onlinecite{fertig99}.

As we have seen, the GVM+replicas method, together with the $e_0$
constraint, yields results that generally compare favorably with experiment.
Other methods such as the SCBA yield much sharper resonances than are
observed. Nevertheless, it is possible that such sharp resonances are the
correct response for the model we are analyzing, in which case the mechanism
that leads to pinning peaks with $Q \sim 1$ in experiment remains unknown.
We leave these questions for future research.

\section{Conclusion}

We have studied the behavior of the pinning peak of the Wigner and bubble
crystals in Landau level $N=2$ as a function of the filling factor. We used
an elastic action obtained from the Hartree-Fock and time-dependent
Hartree-Fock approximations and treated the disorder averaging using the
replica trick and Gaussian Variational Method. Comparisons with recent
microwave experiments in $N=2$ show that the predictions of our model for
the dependence of the pinning frequency and width of the pinning peak with
filling factor compare favorably with experiment. At the moment, however,
our approach can not reproduce the pinning frequencies observed in some
range of filling factor in $N=2$ where the Wigner crystal and the $M=2$
bubble phases are believed to coexist. Finally, we discussed the relation of
this method with those that predict a very narrow resonance.

\section{Acknowledgements}

The authors want to thank R. Lewis, L. Engel and Y. Chen and C. Doiron for
several useful discussions. This work was supported by a research grant (for
R.C.) and undergraduate research grants (for A.F.) both from the Natural
Sciences and Engineering Research Council of Canada (NSERC) and by a
research grant from the Fonds Qu\'{e}b\'{e}cois pour la recherche sur la
nature et les technologies (FQRNT). H.A.F. acknowledges the support of NSF
through Grant No. DMR-0454699.

\appendix

\section{Long wavelength expansion of the dynamical matrix}

The dynamical matrix is given by Eq. (\ref{a_10}) \textit{i.e.}%
\begin{equation}
\overleftrightarrow{D}\left( \mathbf{k}\right) =\frac{2\pi n_{s}e^{2}}{%
\kappa }\sum_{\mathbf{K}}\left[ \Upsilon \left( \mathbf{k+K}\right)
\left\vert \Lambda \left( \mathbf{k+K}\right) \right\vert ^{2}\left( \mathbf{%
k+K}\right) \left( \mathbf{k+K}\right) -\Upsilon \left( \mathbf{K}\right)
\left\vert \Lambda \left( \mathbf{K}\right) \right\vert ^{2}\mathbf{KK}%
\right] ,  \label{d_1}
\end{equation}%
where%
\begin{eqnarray}
\Lambda \left( \mathbf{k}\right) &\equiv &h^{2}\left( \mathbf{k}\right) , \\
\Upsilon \left( \mathbf{k}\right) &\equiv &\frac{1}{\left\vert \mathbf{k+K}%
\right\vert }.
\end{eqnarray}%
We want to find an expression for the dynamical matrix in the
long-wavelength limit valid up to order $k^{3}$. For the form factor, we use
Eqs. (\ref{a_5}),(\ref{a_6}). We expand the interaction and the form factors
to order $k^{3}$ and define in this way the coefficients $\Upsilon
_{0},\Upsilon _{1},\Upsilon _{2},\Upsilon _{3}$ and $\Lambda _{0},\Lambda
_{1},\Lambda _{2},\Lambda _{3}:$ 
\begin{eqnarray}
\Upsilon \left( \left\vert \mathbf{k+K}\right\vert \right) &\approx
&\Upsilon _{0}\left( K\right) +\Upsilon _{1}\left( K\right) \left( \mathbf{%
k\cdot K}\right) +\Upsilon _{2}\left( K\right) k^{2}+\Upsilon _{3}\left(
K\right) \left( \mathbf{k\cdot K}\right) ^{2}, \\
\Lambda \left( \left\vert \mathbf{k+K}\right\vert \right) &\approx &\Lambda
_{0}\left( K\right) +\Lambda _{1}\left( K\right) \left( \mathbf{k\cdot K}%
\right) +\Lambda _{2}\left( K\right) k^{2}+\Lambda _{3}\left( K\right)
\left( \mathbf{k\cdot K}\right) ^{2}.
\end{eqnarray}%
In these expressions, all coefficients depend only on the modulus of $%
\mathbf{K}$. For $\mathbf{K}=0$, we need the expansion to order $k^{2}$ of
the form factor so that we define the coefficients $\lambda _{1}$ and $%
\lambda _{2}$ by

\begin{equation}
\Lambda \left( \mathbf{k}\right) =M^{2}+\lambda _{1}k+\lambda _{2}k^{2}.
\end{equation}%
The coefficients $\lambda _{1},\lambda _{2}$ and $\Lambda _{0},\Lambda
_{1},\Lambda _{2},\Lambda _{3}$ depends on the particular form factor that
is choosen while $\Upsilon _{0}\left( K\right) =\frac{1}{K},\Upsilon
_{1}\left( K\right) =-\frac{1}{K^{3}},$ $\Upsilon _{2}\left( K\right) =-%
\frac{1}{2K^{3}},$ and $\Upsilon _{3}\left( K\right) =\frac{3}{2K^{5}}.$

Using the inversion and reflexion symmetries of the triangular lattice, we
find, after some algebra, that 
\begin{eqnarray}
D_{xx}\left( \mathbf{k}\right) &=&\left( \frac{2\pi n_{s}e^{2}}{\kappa }%
\right) \left[ \left( \frac{M^{2}}{k}+\alpha _{0}+\alpha _{1}+2\alpha
_{2}+\alpha _{3}+\alpha _{4}+\beta _{0}k\right) k_{x}^{2}+\left( \alpha _{3}+%
\frac{\alpha _{4}}{3}\right) k_{y}^{2}\right] , \\
D_{yy}\left( \mathbf{k}\right) &=&\left( \frac{2\pi n_{s}e^{2}}{\kappa }%
\right) \left[ \left( \frac{M^{2}}{k}+\alpha _{0}+\alpha _{1}+2\alpha
_{2}+\alpha _{3}+\alpha _{4}+\beta _{0}k\right) k_{y}^{2}+\left( \alpha _{3}+%
\frac{\alpha _{4}}{3}\right) k_{x}^{2}\right] , \\
D_{xy}\left( \mathbf{k}\right) &=&\left( \frac{2\pi n_{s}e^{2}}{\kappa }%
\right) \left( \frac{M^{2}}{k}+\alpha _{0}+\alpha _{1}+2\alpha _{2}+\frac{%
2\alpha _{4}}{3}+\beta _{0}k\right) k_{x}k_{y}=D_{yx}\left( \mathbf{k}%
\right) ,
\end{eqnarray}%
where we have defined $\alpha _{0}=\lambda _{1},$ $\beta _{0}=\lambda _{2},$
and%
\begin{equation}
\alpha _{1}=\sum_{\mathbf{K\neq 0}}\Lambda _{0}\left( K\right) \Upsilon
\left( K\right) ,
\end{equation}%
\begin{equation}
\alpha _{2}=\sum_{\mathbf{K\neq 0}}\left[ \Upsilon _{0}\left( K\right)
\Lambda _{1}\left( K\right) +\Upsilon _{1}\left( K\right) \Lambda _{0}\left(
K\right) \right] K_{x}^{2},
\end{equation}%
\begin{equation}
\alpha _{3}=\sum_{\mathbf{K\neq 0}}\left[ \Lambda _{0}\left( K\right)
\Upsilon _{2}\left( K\right) +\Lambda _{2}\left( K\right) \Upsilon
_{0}\left( K\right) \right] K_{x}^{2},
\end{equation}%
\begin{equation}
\alpha _{4}=\sum_{\mathbf{K\neq 0}}\left[ \Lambda _{0}\left( K\right)
\Upsilon _{3}\left( K\right) +\Lambda _{1}\left( K\right) \Upsilon
_{1}\left( K\right) +\Lambda _{3}\left( K\right) \Upsilon _{0}\left(
K\right) \right] K_{x}^{4}.
\end{equation}%
With the dynamical matrix written as%
\begin{equation}
D_{\alpha ,\beta }\left( \mathbf{k}\right) =n_{s}^{-1}\left( \lambda +\mu
\right) k_{\alpha }k_{\beta }+n_{s}^{-1}\mu k^{2}\delta _{\alpha ,\beta },
\end{equation}%
the Lam\'{e} coefficients are thus given by%
\begin{eqnarray}
\lambda &=&\left( \frac{2\pi n_{s}^{2}e^{2}}{\kappa }\right) \left( \frac{%
M^{2}}{k}+\alpha _{0}+\alpha _{1}+2\alpha _{2}-\alpha _{3}+\frac{\alpha _{4}%
}{3}+\beta _{0}k\right) , \\
\mu &=&\left( \frac{2\pi n_{s}^{2}e^{2}}{\kappa }\right) \left( \alpha _{3}+%
\frac{\alpha _{4}}{3}\right) .
\end{eqnarray}%
The correction $\beta _{0}k$ to the Lam\'{e} coefficient $\lambda $ (and so
to the bulk modulus) comes from the $\mathbf{K}=0$ term in the summation
over $K$ in Eq. (\ref{d_1}) $i.e.$ it comes from the long-range part of the
Coulomb interaction.

\end{document}